\documentclass[12pt,a4paper]{article}
\usepackage{color,graphics,amsmath,epsfig,rotating,cite}

\begin{document}

\setlength{\unitlength}{1mm}

\def\mbf             {\boldmath}

\def\ti              {\tilde}

\def\a               {\alpha}
\def\b               {\beta}
\def\d               {\delta}
\def\D               {\Delta}
\def\g               {\gamma}
\def\G               {\Gamma}
\def\l               {\lambda}
\def\t               {\theta}
\def\s               {\sigma}
\def\S               {\Sigma}
\def\x               {\chi}

\def\st              {\ti t}
\def\sb              {\ti b}
\def\stau            {\ti \tau}
\def\snu             {\ti \nu}

\def\ch              {\ti \x^\pm}
\def\nt              {\ti \x^0}
\def\sg              {\ti g}
\def\sq              {\ti q}

\def\half            {\frac{1}{2}}
\def\third           {\frac{1}{3}}

\newcommand{\mst}[1]   {m_{\ti t_{#1}}}
\newcommand{\msb}[1]   {m_{\ti b_{#1}}}
\newcommand{\mstau}[1] {m_{\ti\tau_{#1}}}
\newcommand{\mch}[1]   {m_{\ti \x^\pm_{#1}}}
\newcommand{\mnt}[1]   {m_{\ti \x^0_{#1}}}
\newcommand{\msg}      {m_{\ti g}}

\newcommand{\Emiss}      {E_T^{\rm miss}}

\newcommand{\ifb}	{\text{fb}^{-1}}

\newcommand{\change}     {\marginpar{\bf change}}

\newcommand{\gsim}{\;\raisebox{-0.9ex}
           {$\textstyle\stackrel{\textstyle >}{\sim}$}\;}

\newcommand{\lsim}{\;\raisebox{-0.9ex}{$\textstyle\stackrel{\textstyle<}
           {\sim}$}\;}

\newcommand{\smaf}[2] {{\textstyle \frac{#1}{#2} }}


\vspace*{-2cm}
\begin{flushright}
  CERN-PH-TH/2005-266\\ 
  hep-ph/0512284
\end{flushright}
\vspace*{4mm}

\begin{center}
{\Large\bf  Same-sign top quarks\\[2mm]
            as signature of light stops at the CERN LHC} \\[8mm]
{\large     S.~Kraml$^1$ and A.R.~Raklev$^2$}\\[4mm]
{\it 1) Theory Division, Dept.\ of Physics, CERN, CH-1211 Geneva 23, Switzerland\\
     2) Dept.\ of Physics and Technology, University of Bergen, N-5007 Bergen, Norway}\\[8mm]

\end{center}

\begin{abstract}
We present a new method to search for a light scalar top with
$\mst{1}\lsim m_t$, decaying dominantly into a $c$-jet and the lightest
neutralino, at the LHC. The principal idea is to exploit the Majorana
nature of the gluino, leading to same-sign top quarks in events of
gluino-pair production followed by gluino decays into top and
stop. The resulting signature is 2 $b$-jets plus 2 same-sign leptons
plus additional jets and missing energy. We perform a Monte Carlo
simulation for a benchmark scenario, which is in agreement with the
recent WMAP bound on the relic density of dark matter, and demonstrate
that for $\msg\lsim 900$~GeV and $m_{\ti q}>\msg$ the signal can be
extracted from the background. Moreover, we discuss the determination
of the stop and gluino masses from the shape of invariant-mass
distributions. The derivation of the shape formulae is also given.
\end{abstract}

\section{Introduction}

Owing to the large top Yukawa coupling, the supersymmetric partners of the 
top quark, the so-called `scalar tops' or `stops', play a special role in 
the MSSM \cite{Haber:1984rc}. This is manifest in 
i)~the mixing of the left- and right-chiral states $\st_{\rm L,R}$  
to mass eigenstates $\st_{1,2}$ with the mixing and mass splitting 
proportional to $M_{\rm LR}^2 = h_t v_2 (A_t-\mu\cot\b)$;
ii)~the large impact of the stop masses on the light Higgs mass $m_h$ 
through radiative corrections; and 
iii)~the influence of the stop sector on the renormalization group (RG) 
running of the SUSY breaking parameters. 
RG running and L--R mixing can render the lighter stop, $\st_1$, much
lighter than all other squarks. Indeed, there are scenarios which prefer 
a very light $\st_1$, lighter than the top quark. For example, the
requirement of a strong enough first order phase transition to
preserve the baryon asymmetry of the Universe, together with the Higgs
mass bound from LEP2, favour $\mst{1}< m_t$ and somewhat small values of
$\tan\b\sim 2$--8
\cite{Delepine:1996vn,Carena:1997ki,Cline:1998hy,Balazs:2004bu}.
A light $\st_1$ also ameliorates the fine-tuning in SUSY models.

The experimental bound on $\mst{1}$ from LEP2 \cite{lepsusy} is about 
94--100 GeV, depending on the decay mode, the $\st$ mixing, and the 
mass difference between the $\st_1$ and the lightest SUSY particle (LSP). 
This leaves us with 100 GeV $<\mst{1}<m_t$ 
as a very interesting possibility to consider. 
A $\st_1$ in this range may be discovered at the 
Tevatron~\cite{Demina:1999ty,Abel:2000vs} provided the $\st_1$--LSP mass 
difference is large enough. 
At the LHC, studies of scalar tops typically suffer from an enormous 
SUSY background, often from $\sb$ decays, which makes it very difficult 
to extract the $\st$ signal \cite{Dydak:1996ft,Hisano:2002xq,Hisano:2003qu}. 
However, it has to be noted that most of these studies were performed
within the mSUGRA scenario, which imposes relations between stop and
sbottom masses, and also between stops/sbottoms and the rest of the
spectrum. In this paper we follow a different approach, discussing the
phenomenology of a light $\st_1$ in the framework of the general MSSM.

Quite generally, pair production of light stops has a large cross
section at both the Tevatron and the LHC, comparable to the $t\bar t$
cross section. The cross sections at next-to-leading order (NLO) are
given in Table~\ref{tab:stopxs}. If $\mst{1}<\mch{1}+m_b$ and
$\mst{1}-\mnt{1}< m_W$, 
the $\st_1$ may decay dominantly into $c\nt_1$
\cite{Hikasa:1987db}. 
This gives a signature of two $c$-jets plus
missing transverse energy $\Emiss$, which may be extracted at the
Tevatron provided $\mst{1}-\mnt{1}\gsim 30$~GeV and $\mnt{1}\lsim
100$~GeV, for an integrated luminosity of
$4~\ifb$~\cite{Demina:1999ty,Abel:2000vs}.
At the LHC, however, it will be exceedingly difficult to use this 
signature for a discovery. 

In this paper, we therefore propose an alternative signature to search
for a light $\st_1$ at the LHC: For masses up to $\sim$1~TeV, gluinos are 
also copiously pair-produced at the LHC. The cross sections at NLO are
given in Table~\ref{tab:gluinoxs}. If $\msg>\mst{1}+m_t$, gluinos
decay into stops with a large branching ratio.  The important point is
that being Majorana particles, they decay into $t\,\st_1^*$ or $\bar
t\,\st_1^{}$ combinations with equal probability. Pair-produced
gluinos therefore give
\begin{equation}
   \sg\sg\to t\bar t\,\st_1^{}\st_1^*, tt\,\st_1^*\st_1^*,\,
             \bar t\bar t\,\st_1^{}\st_1^{}
\label{eq:gluinochain}
\end{equation}
and hence same-sign top quarks in half of the gluino-to-stop decays.
Let now the $W$ stemming from $t\to bW$ in the same-sign top events decay 
leptonically, and let the $\st_1$ decay into $c\nt_1$. 
This gives a signature of two $b$-jets plus 
two same-sign leptons plus jets plus missing transverse energy,
\begin{equation}
    \tilde g\tilde g\to 
    bb\,l^+l^+\: ({\rm or}\: \bar b\bar b\, l^-l^-) 
    + {\rm jets\:} + \Emiss\,,
\label{eq:bbllsignature}
\end{equation}
which is quite peculiar. As we will show, it will serve to remove most 
of the backgrounds, both SM and SUSY, and may hence be used for discovery 
of a light stop at the LHC. It may moreover be used for determining a 
relationship between the gluino, stop and neutralino masses. Furthermore, 
it might offer a possibility to test the Majorana nature of the gluino 
by comparing the number same-sign and opposite-sign di-tops from 
Eq.~(\ref{eq:gluinochain}). 
The signature Eq.~(\ref{eq:bbllsignature}) will 
therefore be useful for LHC analysis even in the case that 
the stop is discovered at the Tevatron. 
Of course, gluino--squark and squark-pair production with subsequent 
$\ti q\to\sg q$ decay can also contribute to the signal, 
although with additional jets. For the masses of interest in this study, 
$\s(pp\to\sg\ti q)$ is comparable in size to $\s(pp\to\sg\sg)$,
while $\s(pp\to\ti q\ti q^{(*)})$ is smaller by about a factor of five.

\begin{table}[t]\begin{center}
\begin{tabular}{l|rrrrrrr}
  $\mst{1}$ [GeV] & \quad 120 & \quad 130 & \quad 140 & \quad 150 
            & \quad 160 & \quad 170 & \quad 180 \\
\hline
  $\sigma(\st_1^{}\st_1^*)$, Tevatron 
       & 5.43 & 3.44 & 2.25 & 1.50 & 1.02 & 0.71 & 0.50 \\
  $\sigma(\st_1^{}\st_1^*)$, LHC      
       & 757 & 532 & 382 & 280 & 209 & 158 & 121 \\ 
\hline
\end{tabular}\end{center}
\caption{NLO cross sections in pb for $\st_1$ pair production at the 
Tevatron and the LHC, computed with {\tt
Prospino2}~\cite{Beenakker:1996ed}. For the radiative corrections,
$m_{\ti q}=1$~TeV for all squarks apart from $\st_1$, and
$\msg=660$ GeV.
\label{tab:stopxs}}
\end{table}

\begin{table}[t]\begin{center}
\begin{tabular}{l|rrrrrrr}
  $\msg$ [GeV] & \quad 400 & \quad 500 & \quad 600 & \quad 700 
            & \quad 800 & \quad 900 & \quad 1000 \\
\hline
  $\sigma(\sg\sg)$ [pb] & 
  113 & 31.6 & 10.4 & 3.84 & 1.56 & 0.68 & 0.31\\
\hline
\end{tabular}\end{center}
\caption{NLO cross sections in pb for gluino-pair production at the LHC, 
computed with {\tt Prospino2} \cite{Beenakker:1996ed}. 
For the radiative corrections, $\mst{1}=150$ GeV and 
$m_{\ti q\not =\st_1}=1$~TeV has been assumed.
\label{tab:gluinoxs}}
\end{table}

The rest of the paper is organised as follows. We first discuss in
Section~2 constraints from the relic density of dark matter. 
In Section~3 we perform a
case study of a light stop for the LHC: the parameters of our benchmark 
scenario are given in Section~3.1, the Monte Carlo
simulation is explained in Section~3.2, the effective mass scale 
is shown in Section~3.3, the signal isolation is discussed in detail in
Section~3.4, and the determination of masses in Section~3.5. In
Section~4 we then present our conclusions. Finally, the derivation of
the formulae for the invariant-mass distributions 
is given the Appendix.

\section{Constraints from relic density}

Requiring that the lightest SUSY particle (LSP) 
provide the right amount of cold dark matter
\begin{equation}
   0.0945 \leq \Omega h^2 \leq 0.1287
\label{eq:wmap}
\end{equation}
at $2\sigma$ \cite{Bennett:2003bz,Spergel:2003cb} 
puts strong constraints on any SUSY scenario.
In the standard approach, the relic density is $\Omega h^2\propto
1/\langle\sigma v\rangle$, where $\langle\sigma v\rangle$ is the
thermally averaged cross section times the relative velocity of
the LSP pair. This thermally averaged effective annihilation
cross section includes a sum over all (co-)annihilation channels for
the LSP. 
For a neutralino LSP, the value of $\Omega h^2$ hence depends on 
the $\nt_1$ mass and decomposition (i.e.\ on the gaugino-higgsino mixing),
as well as on the properties of all other sparticles that 
contribute to 
the annihilation and co-annihilation processes. 
The main channels are 
i) annihilation into fermion pairs via s-channel Z or Higgs exchange,
ii) annihilation into fermion pairs via t-channel sfermion exchange,
iii) annihilation into $WW$ or $ZZ$ via t-channel exchange of charginos 
or neutralinos, and 
iv) co-annihilation with sparticles which are close in mass to the LSP. 
The value of $\Omega h^2$ depends sensitively on the kinematics of 
the dominating process; in case i) e.g.\ on $m_{A}-2\mnt{1}$, and in 
case iv) on the mass difference $\Delta M$ between the neutralino and 
the co-annihilating sparticle, often the lighter stau or as in our study 
also the light stop. 
As has been shown in \cite{Allanach:2004xn,Belanger:2005jk}, 
a shift in $\Delta M$ of only 1~GeV can induce an ${\cal O}(10\%)$ 
change in $\Omega h^2$.

For a given set of gaugino-higgsino parameters one can hence derive
constraints on [part of] the rest of the spectrum in order to satisfy
the WMAP bound of Eq.~(\ref{eq:wmap}). We illustrate this considering
two scenarios motivated by the results on baryogenesis viable light
stop models of \cite{Balazs:2004ae}, but neglecting CP-violating phases
for simplicity. For computing the neutralino relic density, we use 
the program {\tt micrOMEGAs\,1.3} \cite{Belanger:2001fz}. 
In order not to vary too many parameters, we use a
common mass scale of 250~GeV for all sleptons apart from $\stau_1$,
and a common mass of 1~TeV for all squarks apart from $\st_1$,
assuming $\stau_1\sim\stau_{\rm R}$ and $\st_1\sim\st_{\rm R}$.
Varying $\mst{1}$ and $\mstau{1}$ then means adjusting 
$m_{\ti U}$, $m_{\ti E}$, $A_t$, and $A_\tau$ for fixed 
$m_{\ti Q}$, $m_{\ti L}$, $\mu$ and $\tan\b$. 
For the gaugino masses, we assume the GUT relation 
$M_2 = (g_2/g_1)^2 M_1\simeq 2M_1$.

\subsection{\mbf $M_1=110$ GeV, $\mu=300$ GeV, $\tan\b=7$ \label{sect:point1}}

%
\begin{figure}[t]
  \centerline{\epsfig{file=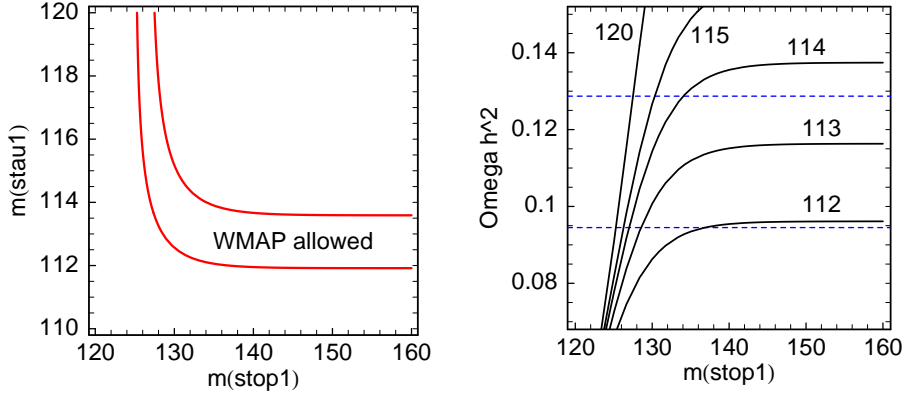,height=6cm}}
\caption{Neutralino relic density for $M_1=110$~GeV, $M_2=220$~GeV, 
$\mu=300$~GeV, $\tan\b=7$: in a) the WMAP allowed band in 
$\mst{1}$--$\mstau{1}$ plane, and in b) $\Omega h^2$ as a function of 
$\mst{1}$ for various $\stau_1$ masses. 
Computed with {\tt micrOMEGAs\,1.3}. All masses are in [GeV]. 
The other parameters are as explained in the text.}
\label{fig:mu300}
\end{figure}

As the first scenario we take the parameter point $M_1=110$~GeV,
$M_2=220$~GeV, $\mu=300$~GeV and $\tan\b=7$.
This gives a spectrum of $\mnt{1...4}=\{105,191,306,340\}$~GeV and
$\mch{1,2}=\{189,340\}$~GeV. The $\nt_1$ is dominantly a bino with
only 1\% wino and 5\% higgsino admixture. With slepton masses around
250~GeV, squark masses around 1~TeV and $m_A\gg 2\mnt{1}$, the $\nt_1$
annihilation cross section is much too low, leading to $\Omega h^2\sim
0.8$, which is well above the WMAP bound. One possibility to achieve
the right $\Omega h^2$ is to lower $m_A$ to about 250~GeV. In this
case the neutralinos annihilate efficiently through $\nt_1\nt_1\to
A\to b\bar b$, leading to $\Omega h^2\sim 0.1$ (when closer to the
pole of $m_A-2\nt_1$, the $\nt$'s annihilate too fast and $\Omega h^2$
becomes too small). Another possibility is to rely on co-annihilation
with stops or staus. This puts rather strong constraints on the
$\stau_1$ and/or $\st_1$ masses, since stau co-annihilation occurs for
$\mstau{1}-\mnt{1}\lsim 10$~GeV while stop co-annihilation requires
$\mst{1}-\mnt{1}\lsim 25$~GeV. These constraints imply that the decay
products of the staus or stops will be difficult to detect at
a hadron collider.

Figure~\ref{fig:mu300}a shows the WMAP-allowed band  
in the $\mst{1}$--$\mstau{1}$ plane for $\st_1\sim\st_{\rm R}$ and 
$\stau_1\sim\stau_{\rm R}$, 
$m_A=1$~TeV and the other parameters as explained above.
Figure~\ref{fig:mu300}b shows the neutralino relic density as a 
function of $\mst{1}$ for various values of $\mstau{1}$. 
As can be seen, the WMAP bound puts a lower limit 
on the $\st_1$ and $\stau_1$ masses of $\mst{1}\gsim 125$~GeV and 
$\mstau{1}\gsim 112$~GeV 
(which is, however, only a stringent bound if one requires 
that the LSP provide all the cold dark matter). 
It can also put an upper limit on one mass as a function of the other. 
This bound is more severe. For example, 
for $\mst{1}\gsim 130$~GeV agreement with WMAP requires 
$\mstau{1}\lsim 114$~GeV if no other process helps the $\nt_1$ 
annihilate efficiently. 
The exact masses of the other sleptons and squarks are not important 
for the computation of the relic density in this example because they are 
too heavy to contribute significantly in (co-)annihilation processes.

\subsection{\mbf $M_1=110$ GeV, $\mu=180$ GeV, $\tan\b=7$}

%
\begin{figure}[t]
  \centerline{\epsfig{file=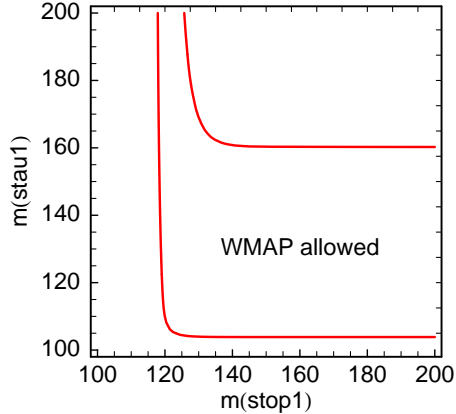,height=6cm}}
\caption{WMAP allowed band in the $\mst{1}$--$\mstau{1}$ plane analogous 
to Fig.~\ref{fig:mu300}a, but for $\mu=180$~GeV.}
\label{fig:mu180}
\end{figure}

For the second scenario, we lower $\mu$ to 180~GeV, keeping the other 
parameters as above. 
This leads to $\mnt{1...4}=\{93,152,188,274\}$~GeV 
and $\mch{1,2}=\{139,273\}$~GeV. The $\nt_1$ has now 4\% wino and  
25\% higgsino admixture and annihilates quite efficently into 
$W^+W^-$, giving a relic density of $\Omega h^2=0.138$. We hence 
need only a small additional contribution to $\langle\sigma v\rangle$,  
e.g.\ from light staus or stops, or from Higgs exchange.  

Figure~\ref{fig:mu180} shows the WMAP allowed band  
in the $\mst{1}$--$\mstau{1}$ plane for this scenario. 
The allowed region is much larger than in Fig.~\ref{fig:mu300}a, 
especially because for $\mst{1}\gsim 130$~GeV, t-channel exchange  
of $\stau_1$ with $\mstau{1}\lsim 160$~GeV is sufficient to bring 
$\Omega h^2$ into the desired range. 
For $\mstau{1}\lsim 105$~GeV, on the other hand, stau co-annihilation 
comes into play, driving $\Omega h^2$ below the $2\sigma$ WMAP bound. 
Likewise, $\Omega h^2$ turns out too low for $\mst{1}\lsim 120$~GeV 
because of co-annihilation with stops. 
Note, however, that in this scenario a $\st_1$ heavier than about 144~GeV 
will decay into $b\ti\x^+_1$ rather than into $c\nt_1$, which gives 
additional $b$ jets and, more importantly, additional same-sign leptons. 
This changes the signature of Eq.~(\ref{eq:bbllsignature}) to 
\begin{equation}
    \sg\sg \to bb\bar b\bar b\,l^+l^+l^-l^- + \Emiss\,,
\label{eq:bbll2signature}
\end{equation}
or more generally to $4b+4W+\Emiss$. It is clear that the information 
from the same-sign top quarks is lost in this way.

\section{Case study for the LHC}

\subsection{Choice of parameters}

In order to explore the feasibility of extracting the light stop
signal Eq.~(\ref{eq:bbllsignature}) 
at the LHC, we perform a case study for the parameters of
Sect.~\ref{sect:point1} ($M_1=110$~GeV, $\mu=300$~GeV, $\tan\b=7$),
$\mst{1}=150$~GeV and $\msg=660$~GeV, assuming BR$(\st_1\to c\nt_1)=1$. 
All squark mass parameters apart
from $m_{\ti U_3}$ are set to 1~TeV in order to maximise BR$(\sg\to
t\st_1)$ and suppress background from e.g.\ $\sb\to t\ti\x^-$
decays. For the sleptons, we take $M_{\ti L,\ti E}=250$~GeV.  As
explained in Sect.~\ref{sect:point1}, agreement with WMAP can be
achieved by $m_A\simeq 250$~GeV or, if $m_A$ is large, by
$\mstau{1}\simeq 112$--113~GeV. In what follows we use
$m_A=250$~GeV. We call this the LST1 scenario. The
parameters and the resulting mass spectrum calculated with {\tt
SuSpect\,2.3} \cite{Djouadi:2002ze} are given in
Tables~\ref{tab:LST1par} and
\ref{tab:LST1mass}, respectively. Note that the SUSY-breaking
parameters in Table~\ref{tab:LST1par} are taken to be on-shell.

\begin{table}
\begin{center}
\begin{tabular}{|c|c|c|c|c|c|c|} \cline{1-5}
$M_1$ & $M_2$ & $M_3$ & $\mu$ & $\tan(\beta)$ \\
110 & 220 & 660 & 300 & 7 \\ \cline{1-5}
$m_A$ & $A_t$ & $A_b$ & $A_\tau$ \\ 
250 & $-670$ & $-500$ & 100 \\ \cline{1-4}
$m_{\tilde L_{1,2}}$ & $m_{\tilde L_3}$ & $m_{\tilde Q_{1,2}}$ & $m_{\tilde Q_3}$ \\
250 & 250  & 1000 & 1000  \\ \cline{1-6}
$m_{\tilde E_{1,2}}$ & $m_{\tilde E_3}$ & $m_{\tilde U_{1,2}}$ & $m_{\tilde D_{1,2}}$ & $m_{\tilde U_3}$  & $m_{\tilde D_3}$ \\   
250 & 250 & 1000 & 1000 & 100 & 1000 \\ \hline
$\alpha_{\mathrm em}^{-1}(m_Z)^{\overline{\mathrm{MS}}}$ & $G_F$ & $\alpha_s(m_Z)^{\overline{\mathrm{MS}}}$ & $m_Z$ & $m_b(m_b)^{\overline{\mathrm{MS}}}$ & $m_t$ & $m_\tau$ \\
127.91 & $1.1664 \times 10^{-5}$ & 0.11720 & 91.187 &  4.2300 & 175.0 & 1.7770 \\ \hline
\end{tabular}
\end{center}
\caption{Input parameters for the LST1 scenario [masses in GeV]. 
Unless stated otherwise, the SM masses are pole masses. 
         \label{tab:LST1par}}
\end{table}

\begin{table}
\begin{center}
\begin{tabular}{|c|c|c|c|c|c|c|c|} \hline
$\tilde d_L$ & $\tilde u_L$ & $\tilde b_1$ & $\tilde t_1$ & $\tilde e_L$ & $ \tilde \tau_1$ & $\tilde \nu_e$ & $\tilde \nu_\tau$\\
1001.69 & 998.60 & 997.43 & 149.63 & 254.35 & 247.00 & 241.90 & 241.90\\ \hline
$\tilde d_R$ & $\tilde u_R$ & $\tilde b_2$ & $\tilde t_2$ & $\tilde e_R$ & $\tilde \tau_2$ \\
1000.30 & 999.40 & 1004.56 & 1019.26 & 253.55 & 260.73 \\ \cline{1-7}
$\tilde g$ & $\tilde \chi^0_1$ & $\tilde \chi^0_2$ & $\tilde \chi^0_3$ & $\tilde \chi^0_4$ & $\tilde \chi^\pm_1$ & $\tilde \chi^\pm_2$ \\
660.00 & 104.81 & 190.45 & 306.06 & 340.80 & 188.64 & 340.09 \\ \cline{1-7}
$h$ & $H$ & $A$ & $H^\pm$ \\
118.05 & 251.52 & 250.00 & 262.45 \\ \cline{1-4}
\end{tabular}
\end{center}
\caption{SUSY mass spectrum [in GeV] for the LST1 scenario. 
For the squarks and sleptons, the first two generations have identical masses.
         \label{tab:LST1mass}}
\end{table}

\subsection{Monte Carlo simulation}

We have generated SUSY events and $t\bar t$ background equivalent to
$30~\text{fb}^{-1}$ of integrated luminosity with {\tt PYTHIA\,6.321}
\cite{PYTHIA} and CTEQ~5L parton distribution functions
\cite{Lai:1999wy}, corresponding to three years running of the LHC
at low luminosity. The SUSY NLO cross sections are found in
Table~\ref{tab:xseccase}. We have also generated additional SM
background in five logarithmic $p_T$ bins from $p_T=50$~GeV to
$4000$~GeV, consisting of $5\times 10^4$ of $W+\rm jet$, $Z+\rm jet$, 
and $WW/WZ/ZZ$ production and $3.5\times 10^5$ QCD 
$2\rightarrow 2$ events per bin.

\begin{table}\begin{center}
\begin{tabular}{c|cccccccc|c}
 & $\s(\st_1\st_1)$ & $\s(\sg\sg)$ & $\s(\sg\sq)$ & $\s(\tilde\chi^0_2 \tilde\chi^\pm_1)$ & $\s(\tilde\chi^\pm_1 \tilde\chi^\mp_1)$ & $\s(\sq\sq)$ & $\s(\sq\sq^*)$ & $\s(\tilde\chi^\pm_1 \tilde g)$ & $\s(t\bar t)$ \\ \hline
LST1 & $280$ & $5.39$ & $4.98$ & $1.48$ & $0.774$ & $0.666$ & $0.281$ & $0.0894$ & $737$\\ \hline
\end{tabular}\end{center}
\caption{The main SUSY NLO cross sections in pb for the LST1 scenario, 
computed with {\sc Prospino2} \cite{Beenakker:1996ed}. 
For comparison, we also give the $t\bar t$ NLO cross section 
taken from \cite{Frixione:2003ei}. 
\label{tab:xseccase}}
\end{table}

Detector simulation has been done using the generic LHC detector simulation\\
{\tt AcerDET\,1.0}~\cite{Richter-Was:2002ch}. This expresses identification
and isolation of leptons and jets in terms of detector coordinates by
azimuthal angle $\phi$, pseudo-rapidity $\eta$ and cone size $\D
R=\sqrt{(\D\phi)^2+(\D\eta)^2}$. We identify a lepton if
$p_T>5(6)$~GeV and $|\eta|<2.5$ for electrons (muons). A lepton is
isolated if it is at a distance $\D R>0.4$ from other leptons and jets, 
and if the transverse energy deposited in a cone $\D R=0.2$ around the
lepton is less than $10$~GeV. Jets are reconstructed by a cone-based
algorithm from clusters and are accepted if the jet has $p_T>15$~GeV
in a cone $\D R=0.4$. The jets are re-calibrated using a flavour
independent parametrisation, optimised to give a proper scale for the
dijet decay of a light (100--120~GeV) Higgs boson. The $b$-tagging efficiency
and light jet rejection is set according to the $p_T$ parametrization
for a low luminosity environment given in \cite{unknown:1997fs}.

\subsection{Effective mass}
\label{sec:meff}
In Fig.~\ref{fig:meff} we show the distribution of the effective mass
$M_{\text{eff}}$, 
\begin{equation}
M_{\text{eff}}=\Emiss+\sum_i p^{\text{jet}}_{T,i},
\end{equation}
for the LST1 scenario under study. The cuts used are:
\begin{itemize}
\item
Require at least four jets with
$p^{\text{jet}}_T>100,100,50,50$~GeV in each event.
\item
No isolated electrons or muons.
\item
$\Emiss>\max(100,0.25\sum p^{\text{jet}}_T)$~GeV.
\end{itemize}
The clear excess of events toward high values of $M_{\text{eff}}$,
compared to the SM distribution, indicates that SUSY should be easily
discovered in our light stop scenario. 

The effective mass has also been shown (see
\cite{Hinchliffe:1996iu,Tovey:2000wk}) to work as a measurement of the
effective SUSY mass scale, $M^{\text{eff}}_{\text{susy}}$, given by
\begin{equation}
M^{\text{eff}}_{\text{susy}}=
M_{\text{susy}}-\frac{m_{\tilde{\chi}^0_1}^2}{M_{\text{susy}}},
\end{equation}
where the SUSY mass scale, $M_{\text{susy}}$, is a cross section
weighted average of the masses of the initially produced SUSY
particles; usually gluinos and squarks. In the light stop scenario,
stop pair production overwhelmingly dominates the SUSY cross
section. Yet it is difficult to extract the stop mass from the
effective mass in this manner, since the cuts placed to handle the SM
background also remove most of the stop pair production events. What
remains is mostly gluino-pair production and gluino-squark production,
and we can instead use the effective mass to estimate the gluino
mass. In \cite{Tovey:2000wk} the linear regression relationship
$M_{\text{est}}=1.7M^{\text{eff}}_{\text{susy}}+134.15$~GeV, where
$M_{\text{est}}$ is the fitted peak of the $M_{\text{eff}}$
distribution, was found from a large set of random mSUGRA and MSSM
models. The joint statistical and systematic error on the SUSY mass
scale from using this relationship was estimated to be below 40\%
after one year running of the LHC at low luminosity. 
With the peak effective mass value of $M_{\text{est}}\simeq 1023$~GeV at LST1, 
and a conservative estimate of accuracy at the level of 30\% after
$30~\ifb$ of integrated luminosity, this gives 
$M^{\text{eff}}_{\text{susy}}= 523\pm 157$~GeV. 
This may serve as a first estimate of the gluino mass scale.

%
\begin{figure}[t]
\centerline{\epsfig{file=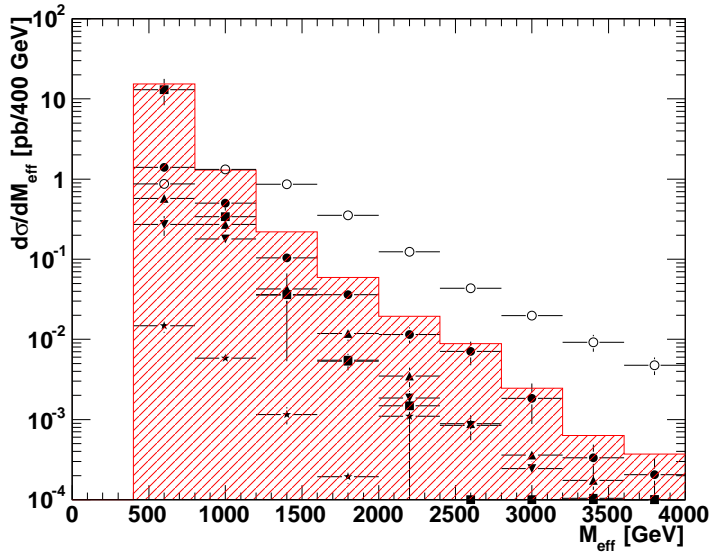,height=8cm}}
\caption{Effective mass distribution. 
SM contributions are $t\bar t$ (filled circles), $W$+jet (triangles), 
$Z$+jet (inverted triangles), $WW/WZ/ZZ$ production (stars) and 
QCD (squares). The sum of all SM events is shown by the hatched 
histogram. SUSY events are shown as open circles.}
\label{fig:meff}
\end{figure}

\subsection{Signal isolation}

\label{sec:signal}
The peculiar nature of the signal with a same-sign top pair will serve
to remove most of the background, both SM and SUSY. We use the
following cuts:\footnote{It should be noted that the cut values have
not been optimized to isolate the signal for this particular model
(LST1); they have been chosen so as to effectively remove SM events
and to isolate events with two same-sign top quarks.}
\begin{itemize}
\item
Require two same sign leptons ($e$ or $\mu$) with $p^{\text{lep}}_T>20$~GeV.
\item
Require at least four jets with $p^{\text{jet}}_T>50$~GeV, at least
two of which are $b$-tagged.
\item
$\Emiss > 100$~GeV.
\item
We cut on top content in the events by requiring two combinations of
leptons and $b$-jets to give invariant masses $m_{bl}<160$~GeV,
consistent with a top.
\end{itemize}

The effects of these cuts are shown in Table~\ref{tab:cut_eff}. The
column ``2lep 4jet'' gives the status after detector simulation and
cuts on two reconstructed and isolated leptons and four reconstructed
jets; ``2$b$'' is the number of events left after the $b$-jet cut,
assuming a $b$-tagging efficiency of $43\%$; ``$\Emiss$'' is the cut
on missing transverse energy and ``SS'' the requirement of two
same-sign leptons. These cuts constitute the signature of
Eq.~(\ref{eq:bbllsignature}). Note the central importance of the same-sign 
cut in removing the SM background, which at that point consists
only of $t\bar t$ events. The cuts on transverse momentum and top
content ``2$t$'' are used to further reduce the background. We find
that the gluino-pair production, followed by gluino decay into top 
and stop and leptonic top decay, is easily
separated from the background.

\begin{table}\begin{center}
\begin{tabular}{l|rrrrrrr}
Cut & 2lep 4jet & $p_T^{\text{lep}}$ & $p_T^{\text{jet}}$ 
    & 2$b$ & $\Emiss$  & 2$t$ & SS\\ \hline
Signal &&&&&&&\\
\ \ $\tilde g \tilde g$ & 10839 & 6317 & 4158 & 960 & 806 & 628 & 330 \\
Background &&&&&&& \\
\ \ SUSY  & 1406 & 778 & 236 & 40 & 33 & 16 & 5 \\
\ \ SM    & 25.3M & 1.3M  & 35977 & 4809 & 1787 & 1653 & 12 \\
\hline
\end{tabular}\end{center}
\caption{Number of events left after cumulative cuts for $30~\ifb$ of integrated luminosity.
\label{tab:cut_eff}}
\end{table}

To investigate other possible backgrounds to our signal we have used
{\tt MadGraph II} with the {{\tt MadEvent} event generator
\cite{Stelzer:1994ta,Maltoni:2002qb}. The search has been limited to parton
level, as we find no processes that can contribute after placing
appropriate cuts. We have looked at SM processes that can mimic a
same-sign top pair by mis-tagging of jets or the production of one or
more additional leptons, as well as inclusive production of same-sign
top pairs. We assume that the two extra jets needed in some cases
could be produced by ISR, FSR, or the underlying event. In particular
we have looked at diffractive scattering $qq \to W^\pm q' W^\pm q'$
and the production of a top pair from gluon radiation in single $W$
production $qq' \to t\bar{t} W^\pm$. Also checked is the production of
$t\bar{t} l^+ l^-$, $t\bar{t}t\bar{t}$, $t\bar{t}t\bar{b}$,
$t\bar{t}b\bar{t}$, $tW^-tW^-$, $\bar{t}W^+\bar{t}W^+$ and $W^\pm
W^\pm b\bar{b}jj$. We place cuts on leptons and quarks as given above,
and demand two lepton-quark pairs consistent with top decays. We also
require neutrinos from the $W$ decays to give the required missing
energy. After these cuts and reasonable detector geometry cuts of $\D
R>0.4$ and $|\eta|<2.5$ for all leptons and quarks, we find that the
cross-sections of all of these processes are too small, by at least an
order of magnitude, to make a contribution at the integrated
luminosity considered.

Last but not least we have assumed that there is no additional same-sign 
top production from flavour-changing neutral currents (FCNC), i.e.\ 
that the anomalous couplings in tgc(u) vertices are effectively zero. 
See \cite{Gouz:1998rk} for a discussion on same-sign tops in FCNC scenarios.

\subsection{Determining masses}

Having isolated the decay chain it will be important to measure the
properties of the sparticles involved to confirm that the decay indeed
involves a light scalar top. Since the neutralino and the neutrino in
the top decay represent missing momentum and energy, reconstruction of
a mass peak is impossible. The well studied alternative to this, see
e.g. \cite{Hinchliffe:1996iu,Bachacou:1999zb,
Allanach:2000kt,Lester:2001zx,Gjelsten:2004}, is to use the invariant-mass
distributions of the SM decay products. Their endpoints can be given
in terms of the SUSY masses involved, and these equations can then in
principle be solved to give the masses.

In our scenario there are two main difficulties with this. First,
there are four possible endpoints: $m_{bl}^{\max}$, $m_{bc}^{\max}$,
$m_{lc}^{\max}$ and $m_{blc}^{\max}$, of which the first simply gives
a relationship between the masses of the $W$ and the top, and the
second and third are linearly dependent, so that we are left with
three unknown masses and only two equations. Second, because of the
information lost with the escaping neutrino the distributions of
interest all fall very gradually to zero. Determining exact endpoints
in the presence of background, while taking into account smearing from
the detector, effects of particle widths etc. will be very difficult.
The shapes of the invariant-mass distributions are shown, for some
arbitrary normalization, in Fig.~\ref{fig:shapes}.

%
\begin{figure}[t]
\centerline{\epsfig{file=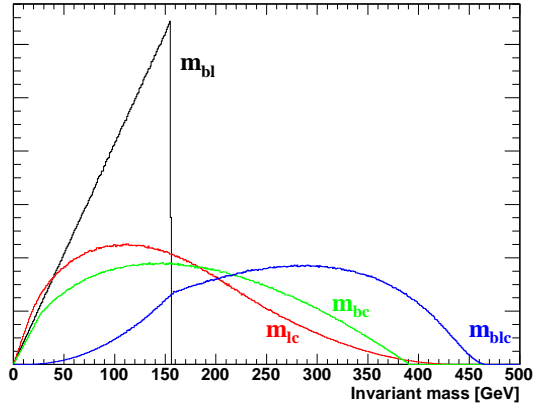,height=6cm}}
\caption{Invariant-mass distributions for LST1. These distributions only take into account the kinematics of the decay, i.e. no spin or width effects are included.}
\label{fig:shapes}
\end{figure}

In what follows we have partially solved the second problem by
extending the endpoint method and deriving analytic expressions for
the shape of the invariant-mass distributions $m_{bc}$ and
$m_{lc}$. The derivation is given in
Appendix~\ref{app:derivation}. These shapes can now be used to fit the
whole distributions of the isolated signal events and not just the
endpoints. This greatly reduces the uncertainty involved in endpoint
determination, and provides the possibility of getting more
information on the masses. One could also imagine extending this
method to include spin effects in the distribution, to get a handle on
the spins of the SUSY particles involved and possibly confirming the
scalar nature of the stop\footnote{For more details on the derivation
of invariant-mass distributions in cascade decays, and the inclusion
of spin effects, see
\cite{Miller:2005zp}.}.

In fitting the $m_{bc}$ and $m_{lc}$ distributions, we start from the
isolated $\tilde g \tilde g$ signal of Section~\ref{sec:signal}.
However, these events contain some where one or both of the $W$ decay
to a tau, which in turn decays leptonically. These taus are an
additional, irreducible background to our distributions. The $b$-jets
and leptons are paired through the cut on two $t$ quark candidates. A
comparison with Monte Carlo truth information from the event
generation shows that this works well in picking the right pairs. The
issue which remains is to identify the $c$-quark initiated jets and to
assign these to the correct $b$-jet and lepton pair. The precision of
our mass determination is limited by systematics from these problems.

Two different strategies can be used for picking the $c$-jets. The
strong correlation between the tagging of $b$- and $c$-jets suggest an
inclusive $b/c$-jet tagging of at least four jets per event. The two
types of jets can then be separated on their $b$-tagging likelihoods,
and the requirement of two top candidates in the event. A thorough
investigation of this strategy requires a full simulation study, using
realistic $b$-tagging routines. The other strategy, which we follow
here, is to accept a low $b$-tagging efficiency to pick two $b$-jets
and reject most $c$-jets. The likelihoods in the $b$-tagging routine
can then help pick the correct $c$-jets from the remaining jets. In
this fast simulation study we are restricted to a simple statistical
model of the efficiency of making this identification. We have looked at
two cases: One worst case scenario with no direct $c$-jet
identification, where we only use the kinematics of the event to pick
out the $c$-jets, and one where we have assumed an additional $20\%$
probability of identifying a $c$-jet directly from the $b$-tagging
likelihood.

For events where we have missed one or both $c$-jets, they are picked
as the two hardest remaining jets with $p_T^{\text{jet}}<100$~GeV. The
upper bound on transverse momentum is applied because the stop is
expected to be relatively light if our signal exists, and it avoids
picking jets from the decay of heavy squarks. Our $c$-jet candidates
are paired to the top candidates by their angular separation in the
lab frame, and by requiring consistency with the endpoints of the two
invariant-mass distributions we are not looking at. E.g. if we wish to
construct the $m_{bc}$ distribution we demand consistency with the
endpoints $m_{lc}^{\max}$ and $m_{blc}^{\max}$
\footnote{We require that the values are below the rough estimates
$m_{bc}^{\max}=430$~GeV, $m_{lc}^{\max}=480$~GeV and
$m_{blc}^{\max}=505$~GeV, approximately 40 GeV above the nominal
values, so no precise pre-determination of endpoints is
assumed.}. Events with no consistent combinations of $c$-jets and top
candidates are rejected.

The fit functions for $m_{bc}$ and $m_{lc}$ are given in
Eqs.~(\ref{eq:m_bc_dist}) and (\ref{eq:m_lc_distA}). In principle both
of the two linearly independent parameters $m_{bc}^{\max}$ and $a$
could be determined by fits. However, we typically have
$m_tm_{\tilde{t}_1}\ll m_{\tilde g}^2$ for light stops, so that $a\approx 1$. 
For LST1, the nominal value is $a=0.991$. The distributions are sensitive 
to such values of $a$ only at very low invariant masses. Because of the 
low number of events, no sensible value can be determined from a fit; 
we therefore set $a=1$. The fit quality and value of $m_{bc}^{\max}$ is 
found to be insensitive to the choice of $a$ for $a\gsim 0.980$.

The results of the fits to $m_{bc}^{\max}$, assuming no $c$-jet
tagging, are shown in Fig.~\ref{fig:im_noctag}. The combined result of
the two distributions is $m_{bc}^{\max}=383.2\pm4.9$~GeV, to be
compared with the nominal value of $391.1$~GeV. The large $\chi^2$
values of the fits and the low value of $m_{bc}^{\max}$ indicate that
there are some significant systematical errors. Comparing to Monte
Carlo truth information we have found that the peaks at around
$100-150$~GeV, responsible for the bad fit quality, are chiefly the
result of events with one or more taus (see above). In
Fig.~\ref{fig:im_ctag} we show the results assuming $20\%$ $c$-tagging
efficiency. The combined result has improved to
$m_{bc}^{\max}=389.8\pm5.3$~GeV. We expect to be able to do better
than this with full information from the $b$-tagging routine.

%
\begin{figure}[t!]
\centerline{\epsfig{file=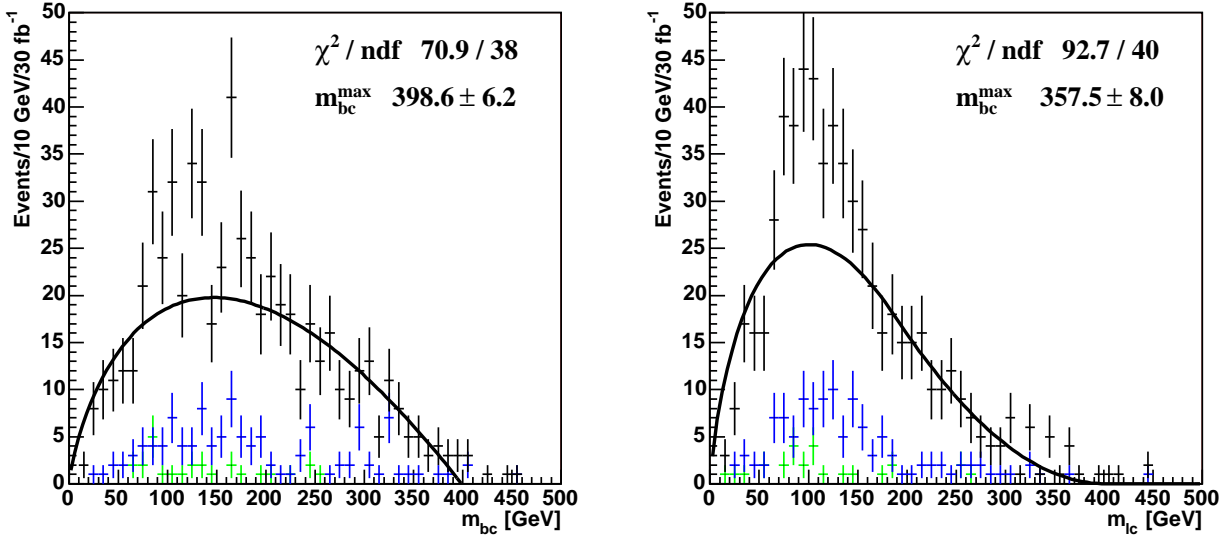,height=8cm}}
\caption{Invariant-mass distributions without $c$-jet tagging 
(black with error bars) and best fit. Left panel shows $m_{bc}$, right
panel $m_{lc}$. Also shown are the contributions from the SM
background (green) and the SUSY background (blue). The SUSY background
consists mostly of events with one or more taus (see text).}
\label{fig:im_noctag}
\end{figure}

%
\begin{figure}[t!]
\centerline{\epsfig{file=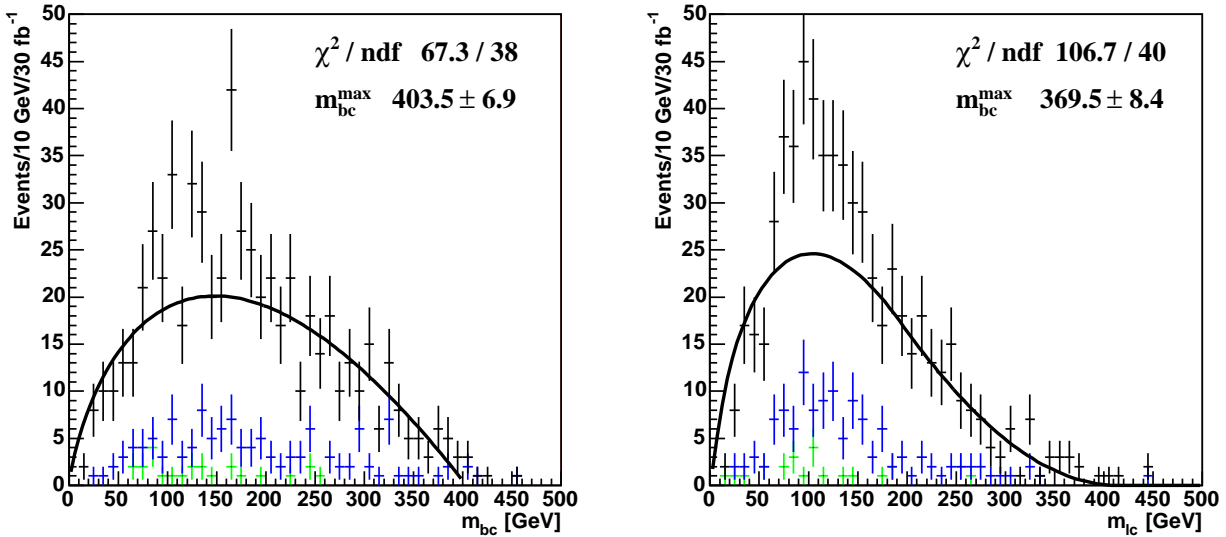,height=8cm}}
\caption{Invariant-mass distributions with $20\%$ $c$-tagging 
efficiency after $b$-tagging. For details see caption of 
Figure~\ref{fig:im_noctag}.}
\label{fig:im_ctag}
\end{figure}

\section{Conclusions}

We have investigated a baryogenesis-motivated scenario of a light stop
($m_{\tilde t_1} \lsim m_t$), with $\tilde t_1\to c\tilde\chi^0_1$ as
the dominant decay mode. In this scenario, pair production of $\tilde
t_1$ leads to a signature of two jets and missing transverse energy,
which is all but promising for the discovery 
of $\tilde t_1$ at the LHC.

We have hence proposed a method which uses instead stops stemming 
from gluino decays: in gluino pair production, the Majorana nature of the 
gluino leads to a peculiar signature of same-sign top quarks in half 
of the gluino-to-stop decays. 
For the case that all other squarks are heavier than the gluino, we have 
shown that the resulting signature of $2b$'s + 2 same-sign leptons + jets 
$+\Emiss$ can be extracted from the background and serve 
as a discovery channel for a light $\tilde t_1$. 

We have also demonstrated the measurement of a relationship between
the gluino, stop and LSP masses. Taken together with a determination
of other invariant-mass endpoints, and a measurement of the SUSY mass
scale from the effective mass scale of events, this may be sufficient
to determine the masses of the SUSY particles involved. 
In particular, if the invariant-mass distributions of the
isolated events fit the predicted shapes, this strengthens the
interpretation of the events as gluino decays into a top and a stop.

Last but not least some comments are in order on the robustness of the
signal. While our analysis has been done for $\msg=660$~GeV, we have
checked that the signal remains significant enough for a 5$\sigma$
discovery for gluino masses up to $\msg\sim$ 900~GeV. In the high-mass
region, gluino-squark production with subsequent squark decay to
gluino is responsible for a significant fraction of events. On the
other hand, if sbottoms are lighter than assumed for LST1, they will
contribute to the SUSY background through $\sb_i\to t\ch_j$
($i,j=1,2$) decays. For instance, lowering the sbottom masses by a
factor of 2, while keeping all other parameters as in LST1, would
increase the background after cuts by 5 events. While this is not
dramatic, the fits to invariant-mass distributions are worse because
of the opening up of new gluino decays $\sg\to b\sb_i$. Lowering the
stop mass while keeping the LSP mass fixed, entering the
stop-coannihilation region, should also make the signal more difficult
to find because the $c$-jets become too soft. Still, we find only a
12\% decrease in the signal when setting $m_{\st_{1}}=120$~GeV. This
can be explained by a large number of signal events passing the cuts
because of squark decays to gluinos producing an extra jet. This
implies that the same-sign tops can be used to search for a light stop
even in the stop-coannihilation channel, unreachable at the
Tevatron. The fit to the invariant-mass shapes is again worse than for
LST1.

We conclude that if $\st_1$ decays into $c\nt_1$, light stops may be 
discovered through the search for
same-sign tops in the decay of pairs of gluinos for a wide range of
SUSY masses. A relation between the gluino, stop and LSP masses
can be determined from invariant-mass distributions.
In this paper we have demonstrated the feasibility of our method; 
a full simulation will clearly be necessary to assess its full potential.

\section*{Acknowledgements}
We thank Tilman Plehn and Tim Stelzer for helpful discussions on 
{\tt MadGraph}. 
S.K.\ is supported by an APART (Austrian Programme of Advanced
Research and Technology) grant of the Austrian Academy of
Sciences. A.R.R.\ is supported by the Norwegian Research Council.

\appendix

\section{Derivation of shape formulae}
\label{app:derivation}

The differential decay width for two independent angular variables
(cosines) $u$ and $v$, in a decay with a given spin configuration, is
\begin{equation}
\frac{1}{\Gamma}\frac{\partial^2\Gamma}{\partial u\partial v}
=f_u(u)\theta(u-1)\theta(1-u)f_v(v)\theta(v-1)\theta(1-v),
\end{equation}
where the spin information is contained in the functions $f_u(u)$ and
$f_v(v)$ and the theta functions are ordinary step function, limiting
the values of $u$ and $v$ that give non-zero decay width.

By a change of variables one can easily go from an expression of the
invariant-mass in terms of these variables, to the distribution sought. Given the invariant-mass $m(u,v)$, we have
\begin{eqnarray}
\frac{1}{\Gamma}\frac{\partial\Gamma}{\partial m^2}
& = & \int_{-\infty}^{\infty}\frac{1}{\Gamma}\frac{\partial^2\Gamma}{\partial m^2\partial v}dv \nonumber \\
& = & \int_{-\infty}^{\infty}\left|\frac{\partial\left(u,v\right)}{\partial\left(m^2,v\right)}\right|\frac{1}{\Gamma}\frac{\partial^2\Gamma}{\partial u\partial v}dv. \label{eq:d_width}
\end{eqnarray}
An extension to three or more independent angular variables is trivial.

We now proceed to derive the shape of the distribution of the
invariant-masse of the $b$- and the $c$-quark, $m_{bc}$, and the
invariant-mass of the lepton and the $c$-quark, $m_{lc}$. We will assume
that all particles are spin-0, so that we have isotropic decays in
particle rest frames. This lets us set $f_u$ and $f_v$ to be
constant. This result is what we expect when summing over all final
states. Effects of non-zero spin can be introduced via these functions.
We also assume that the lighter quarks, $b$ and $c$, are massless.

\subsection{$m_{bc}$}
In the rest frame (RF) of the stop, energy and momentum conservation in
the decay of the gluino, stop and top gives
\begin{eqnarray}
E_c^{\tilde{t}_1} & = & \frac{m_{\tilde{t}_1}^2-m_{\tilde{\chi}_1^0}^2}{2m_{\tilde{t}_1}},
\label{eq:E_c_t1}\\
E_t^{\tilde{t}_1} & = & \frac{m_{\tilde{g}}^2-m_{t}^2-m_{\tilde{t}_{1}}^2}{2m_{\tilde{t}_1}},
\label{eq:E_t_t1}\\
E_b^{\tilde{t}_1} & = &\frac{m_{t}^{2}-m_{W}^{2}}{2E_t^{\tilde{t}_1}-2\sqrt{(E_t^{\tilde{t}_1})^2-m_t^2}\cos\theta_{tb}^{\tilde{t}_1}},
\label{eq:E_b_t1}
\end{eqnarray}
where $E_c^{\tilde{t}_1}$, $E_t^{\tilde{t}_1}$ and $E_b^{\tilde{t}_1}$
are the energies of $c$, $t$ and $b$ respectively in the stop RF and
$\theta_{tb}^{\tilde{t}_1}$ is the angle between $t$ and $b$ in that frame.

We can rewrite $\cos\theta_{tb}^{\tilde{t}_1}$ in terms of an
isotropically distributed angle, $\theta_{\tilde{g}b}^t$, in the top
RF.\footnote{We here make the assumption that the top has
spin-0. Again this is what a sum over final states will yield.}  By
expressing $m_{\tilde{g}b}^2$ in the two rest frames, and since
$\cos\theta_{tb}^{\tilde{t}_1}=\cos\theta_{\tilde{g}b}^{\tilde{t}_1}$,
we get
\begin{equation}
E_b^{\tilde{t}_1}(E_{\tilde{g}}^{\tilde{t}_1}-\sqrt{(E_{\tilde{g}}^{\tilde{t}_1})^2-m_{\tilde{g}}^2}\cos\theta _{tb}^{\tilde{t}_1})
=E_b^t(E_{\tilde{g}}^t-\sqrt{(E_{\tilde{g}}^t)^2-m_{\tilde{g}}^2}\cos\theta_{\tilde{g}b}^t).
\end{equation}
In the top RF,  from the conservation of energy and momentum in
the decays of the gluino and the top, we have that:
\begin{eqnarray}
E_b^t & = &\frac{m_{t}^2-m_{W}^2}{2m_{t}},
\label{eq:E_b_t}\\
E_{\tilde{g}}^t & = & \frac{m_{\tilde{g}}^2+m_t^2-m_{\tilde{t}_1}^2}{2m_t}.
\label{eq:E_g_t}
\end{eqnarray}
Solving for $\cos\theta _{tb}^{\tilde{t}_1}$, using (\ref{eq:E_t_t1}),
(\ref{eq:E_b_t1}), (\ref{eq:E_b_t}) and (\ref{eq:E_g_t}), we then
arrive at
\begin{equation}
\cos\theta_{tb}^{\tilde{t}_1}=\frac{a-\cos\theta_{\tilde{g}b}^t}
{1-a\cos\theta_{\tilde{g}b}^t}
\label{eq:t_tb_t1}
\end{equation}
where $a$ is given by
\begin{equation}
a=\frac{m_2^2}{m_1^2},
\label{eq:a}
\end{equation}
with $m_1$ and $m_2$ defined as
\begin{eqnarray}
m_1^2 & = & m_{\widetilde{g}}^2-m_{t}^2-m_{\widetilde{t}_1}^2\label{eq:m1},\\
m_2^4 & = & m_{1}^4-4m_{t}^2m_{\widetilde{t}_{1}}^2. \label{eq:m2}
\end{eqnarray}
We can now find an expression for $m_{bc}^2:$
\begin{eqnarray}
m_{bc}^2 & = & 2E_b^{\tilde{t}_1}E_c^{\tilde{t}_1}\left(1-\cos\theta _{bc}^{\tilde{t}_{1}}\right) \nonumber \\
& = & \left(m_{bc}^{\max}\right)^2\left(\frac{1-a\cos\theta_{\tilde{g}b}^{t}}{1+a}\right)\left(\frac{1-\cos\theta_{bc}^{\tilde{t}_1}}{2}\right),
\end{eqnarray}
where the endpoint of the distribution, $m_{bc}^{\max}$, is
\begin{equation}
\left(m_{bc}^{\max}\right)^2=\frac{\left(m_{t}^2-m_{W}^2\right)}{m_{t}^2}\frac{\left(m_{\widetilde{t}_{1}}^2-m_{\widetilde{\chi}_{1}^{0}}^2\right)\left(m_{1}^2+m_{2}^2\right)}{2m_{\widetilde{t}_{1}}^2}.
\end{equation}

Using the switch of variables 
\begin{equation}
u = \frac{1-\cos\theta_{bc}^{\tilde{t}_1}}{2}, \quad\quad v = \frac{1-\cos\theta_{\tilde{g}b}^t}{2},
\end{equation}
we can write the invariant-mass as
\begin{equation}
m_{bc}^2 = \left(m_{bc}^{\max}\right)^2\left(\frac{1-a+2av}{1+a}\right)u.
\end{equation}
From Eq.~(\ref{eq:d_width}), and setting $f_u(u)=f_v(v)=1$, we find the
distribution of the invariant-mass
\begin{eqnarray}
\frac{1}{\Gamma }\frac{\partial \Gamma }{\partial m_{bc}^{2}}
&=&\int_{-\infty }^{\infty }\left| \frac{\partial \left( u,v\right) }{%
\partial \left( m_{bc}^{2},v\right) }\right| \theta \left( u\right) \theta
\left(1-u\right)\theta\left(v\right)\theta\left(1-v\right)dv \nonumber \\
&=&\frac{1+a}{\left( m_{bc}^{\max }\right) ^{2}}\int_{0}^{1}%
\frac{1}{1-a+2av}\,\theta\left(1-\frac{m_{bc}^{2}\left(1+a\right)}{\left( m_{bc}^{\max}\right)^{2}\left(1-a+2av\right)}\right)dv.
\end{eqnarray}
The theta function leads to two different lower limits on the integration, depending on the value of $m_{bc}$. Performing the integration gives
\begin{equation}
\frac{1}{\Gamma }\frac{\partial \Gamma }{\partial m_{bc}^{2}}=\left\{ 
\begin{array}{cl}
\displaystyle\frac{1+a}{2a(m_{bc}^{\max})^2}\ln\frac{1+a}{1-a}\quad
&\text{for~~}0<m_{bc}^2<(m_{bc}^{\max})^2\frac{1-a}{1+a}, \\ 
\displaystyle\frac{1+a}{2a\left(m_{bc}^{\max}\right)^2}\ln\frac{(m_{bc}^{\max})^2}{m_{bc}^2}\quad
&\text{for~~}\left(m_{bc}^{\max}\right)^2\frac{1-a}{1+a}<m_{bc}^2<(m_{bc}^{\max})^2,
\end{array}
\right.
\label{eq:m_bc_dist}
\end{equation}
expressing the distribution in terms of two parameters, $a$ and
$m_{bc}^{\max}$.

\subsection{$m_{lc}$}
In the top RF, energy and momentum conservation in the decay of the
gluino, stop, top and $W$ gives, in addition to Eq.~(\ref{eq:E_b_t}),
\begin{eqnarray}
E_l^t & = &\frac{m_tm_W^2}{(m_t^2+m_W^2)-(m_t^2-m_W^2)\cos\theta_{Wl}^t},
\label{eq:E_l_t} \\
E_c^t & = & \frac{m_t(m_{\tilde{t}_1}^2-m_{\tilde{\chi}_1^0}^2)}{m_1^2-m_2^2\cos\theta_{\tilde{t}_{1}c}^t},
\label{eq:E_c_t}
\end{eqnarray}
where $E_l^t$ and $E_c^t$ are the energies of $b$, $l$ and
$c$ respectively in the top RF, $\theta_{Wl}^t$ is the angle between
$W$ and $l$ in that frame and $\theta_{\tilde{t}_1c}^t$ is the angle
between $\tilde{t}_1$ and $c$.

We again change angles to isotropically distributed angles (under spin-0
assumptions). From $\cos\theta_{Wl}^t=-\cos\theta_{bl}^t$, and expressing
$m_{bl}^{2}$ in both the top and $W$ rest frames, we have
\begin{equation}
E_b^tE_l^t(1+\cos\theta_{Wl}^t)=E_b^WE_l^W(1-\cos\theta_{bl}^W).
\end{equation}
In the $W$ RF, energy and momentum conservation in the top and $W$ decays gives
\begin{eqnarray}
E_b^W & = & \frac{m_t^2-m_W^2}{2m_W},
\label{eq:E_b_W} \\
E_l^W & = & \half m_W.
\label{eq:E_l_W}
\end{eqnarray}
Using (\ref{eq:E_b_t}), (\ref{eq:E_c_t}), (\ref{eq:E_b_W}) and
(\ref{eq:E_l_W}) we find that
\begin{equation}
\cos\theta_{Wl}^t=\frac{(m_t^2-m_W^2)-(m_t^2+m_W^2)\cos\theta_{bl}^W}{(m_t^2+m_W^2)-(m_t^2-m_W^2)\cos\theta_{bl}^W}.
\end{equation}

From $\cos\theta_{\tilde{t}_{1}c}^t=\cos\theta_{\tilde{g}c}^t$, and
expressing $m_{\tilde{g}c}^2$ in both the $t$ and $\st_{1}$ rest frames we
get
\begin{equation}
E_c^t(E_{\tilde{g}}^t-\sqrt{(E_{\tilde{g}}^t)^2-m_{\tilde{g}}^2}\cos\theta_{\tilde{t}_{1}c}^t)=
E_c^{\tilde{t}_1}(E_{\tilde{g}}^{\tilde{t}_1}-\sqrt{(E_{\tilde{g}}^{\tilde{t}_1})^2-m_{\tilde{g}}^2}\cos\theta_{\tilde{g}c}^{\tilde{t}_1})
\end{equation}
from which, using (\ref{eq:E_c_t1}), (\ref{eq:E_t_t1}),
(\ref{eq:E_g_t}) and (\ref{eq:E_c_t}), it follows that
\begin{equation}
\cos\theta_{\tilde{t}_{1}c}^t=\frac{a-\cos\theta_{\tilde{g}c}^{\tilde{t}_1}}{1-a\cos\theta_{\tilde{g}c}^{\tilde{t}_1}}.
\end{equation}
This could also have been found from (\ref{eq:t_tb_t1}), using a
symmetry argument.

We can now find an expression for $m_{lc}^2:$
\begin{eqnarray}
m_{lc}^2 & = & 2E_l^tE_c^t\left(1-\cos\theta_{lc}^t\right) \nonumber \\
& = & (m_{lc}^{\max})^2
\left(\frac{1-a\cos\theta_{\tilde{g}c}^{\tilde{t}_{1}}}{1+a}\right)
\left(\frac{m_t^2+m_W^2-\left(m_t^2-m_W^2\right)\cos\theta_{bl}^W}{2m_t^2}\right)
\left(\frac{1-\cos\theta_{lc}^t}{2}\right),
\end{eqnarray}
where 
\begin{equation}
(m_{lc}^{\max })^2 = \frac{\left(m_{\tilde{t}_1}^2-m_{\tilde{\chi}_1^0}^2\right)\left(m_1^2+m_2^2\right)}{2m_{\tilde{t}_1}^2}.
\end{equation}

Using the switch of variables 
\begin{equation}
u = \frac{1-\cos\theta_{\tilde{g}c}^{\tilde{t}_1}}{2}, \quad\quad
v = \frac{1-\cos\theta_{bl}^W}{2}, \quad\quad
w = \frac{1-\cos\theta_{lc}^t}{2},
\end{equation}
we can write 
\begin{equation}
m_{lc}^2 = (m_{lc}^{\max})^2
\left(\frac{1-a+2au}{1+a}\right)
\left(\frac{m_W^2+\left(m_t^2-m_W^2\right)v}{m_t^2}\right)w.
\end{equation}
Since
\begin{eqnarray}
\frac{1}{\Gamma }\frac{\partial^3\Gamma}{\partial u\partial v\partial
m_{lc}^2}
& = & \frac{1}{\Gamma}\frac{\partial^3\Gamma}{\partial
u\partial v\partial w}\left|\frac{\partial\left(u,v,w\right)}{\partial
\left(u,v,m_{lc}^2\right)}\right| \nonumber \\
& = & \frac{m_{t}^{2}(1+a)\hat{\theta}(u)\hat{\theta}(v)\hat{\theta}(w)}{(m_{lc}^{\max})^2\left(1-a+2au\right)\left(m_W^2+\left(m_t^2-m_W^2\right)v\right)},
\end{eqnarray}
where $\hat{\theta}(x)=\theta(x)\theta(1-x)$, we have from a
generalization of (\ref{eq:d_width}) that
\begin{equation}
\frac{1}{\Gamma}\frac{\partial\Gamma}{m_{lc}^2}
= \frac{m_t^2(1+a)}{(m_{lc}^{\max})^2}\int_{0}^{1}\int_{0}^{1}\frac{\hat{\theta}(w)}{(1-a+2au)(m_W^2+(m_t^2-m_W^2)v)}dudv.
\end{equation}
The step functions for $w$ give more complicated bounds on the
integration than what was the case for $m_{bc}$. We must have
$m_{lc}^{2}>0$ and
\begin{equation}
m_W^2+(m_t^2-m_W^2)v>\frac{m_t^2(1+a)m_{lc}^2}{(m_{lc}^{\max})^2(1-a+2au)}.
\end{equation}
This integration and the integration limits from the last inequality are best
explored by yet another change of variables 
\begin{equation}
x = 1-a+2au, \quad y = m_W^2+(m_t^2-m_W^2)v,
\end{equation}
from which the integral can be written 
\begin{equation}
\frac{1}{\Gamma}\frac{\partial\Gamma}{m_{lc}^2}=\frac{1+a}{2a(m_{bc}^{\max})^2}\int_{m_W^2}^{m_t^2}\int_{1-a}^{1+a}\hat{\theta}\left(\frac{m_{lc}^2}{\left(m_{lc}^{\max}\right)^2}\frac{m_t^2(1+a)}{xy}\right)\frac{1}{xy}\,dxdy.
\label{eq:m_lc_int}
\end{equation}

We note that there is a maximum of $m_{lc}^2<(m_{lc}^{\max })^2$, as
expected. There are then five different possible shapes
for the areas of integration in the $xy-$plane. For ease of notation
we use
\begin{equation}
x_1=1-a, \quad x_2=1+a, \quad y_1=m_W^2, \quad y_2=m_t^2.
\end{equation}
The shapes can be categorized by bounds on $y(x_1)$ and $y(x_2)$ as
found in Table~\ref{tab:shape}.
\begin{table}
\begin{center}
\begin{tabular}{|c|c|c|} \hline
Shape & $y(x_1)$         & $y(x_2)$          \\ \hline
I     & $y_2<y(x_1)$     & $y_1<y(x_2)<y_2$  \\ 
II    & $y_1<y(x_1)<y_2$ & $y_1<y(x_2)<y_2$  \\ 
III   & $y_2<y(x_1) $    & $y(x_2)<y_1$      \\ 
IV    & $y_1<y(x_1)<y_2$ & $y(x_2)<y_1$      \\ 
V     & $y(x_1)<y_1$     & $y(x_2)<y_1$      \\ \hline
\end{tabular}
\caption{Shapes for the area of integration for the integral in Eq.~(\ref{eq:m_lc_int}).}
\label{tab:shape}
\end{center}
\end{table}
The inequalities in the categorization of the shapes can be expressed
as bounds on $m_{lc}^2$:
\begin{eqnarray}
y_1 & < & y(x_1)\Leftrightarrow(m_{lc}^{\max})^2\frac{(1-a)}{(1+a)}\frac{m_W^2}{m_t^2}<m_{lc}^2, \\
y_1 & < & y(x_2)\Leftrightarrow(m_{lc}^{\max})^2\frac{m_W^2}{m_t^2}<m_{lc}^2, \\
y_2 & < & y(x_1)\Leftrightarrow(m_{lc}^{\max})^2\frac{(1-a)}{(1+a)}<m_{lc}^2, \\
y_2 & < & y(x_2)\Leftrightarrow(m_{lc}^{\max})^2<m_{lc}^2.
\end{eqnarray}
We can see that there will be two cases, depending on the mass hierarchy: 
\begin{equation}
\begin{tabular}{ll}
Case A: & $\frac{m_W^2}{m_t^2}<\frac{1-a}{1+a}$ \vspace{2mm}\\ 
Case B: & $\frac{m_W^2}{m_t^2}>\frac{1-a}{1+a}$
\end{tabular}
\end{equation}
We give the invariant-mass distribution for both of these cases.

\paragraph{Case A}
Here Shape III is excluded, so the invariant-mass distribution is
\begin{equation}
\frac{1}{\Gamma}\frac{\partial\Gamma}{m_{lc}^2}= \left\{ 
\begin{array}{lcl}
\displaystyle\frac{1+a}{2a(m_{bc}^{\max})^2}\ln\frac{1+a}{1-a}\ln\frac{m_t^2}{m_W^2}
& \text{for~}
& 0<m_{lc}^{2}<(m_{lc}^{\max})^2\frac{1-a}{1+a}\frac{m_W^2}{m_t^2}, \\[5mm]
\lefteqn{\frac{1+a}{2a(m_{bc}^{\max})^2}\left[\ln\frac{1+a}{1-a}\ln\frac{m_t^2}{m_W^2}-\frac{1}{2}\left(\ln\frac{1+a}{1-a}\frac{m_t^2}{m_W^2}\frac{m_{lc}^2}{(m_{lc}^{\max})^2}\right)^2\right]}   \hspace{30mm} \\[5mm]
& \text{for~}
& (m_{lc}^{\max})^2\frac{1-a}{1+a}\frac{m_W^2}{m_t^2}<m_{lc}^2<(m_{lc}^{\max})^2\frac{m_W^2}{m_t^2}, \\[5mm]
\lefteqn{\frac{1+a}{2a(m_{bc}^{\max})^2}\ln\frac{1+a}{1-a}\left(\ln\frac{\left(m_{lc}^{\max}\right)^2}{m_{lc}^2}-\frac{1}{2}\ln\frac{1+a}{1-a}\right)} \\[5mm]
& \text{for~}
& (m_{lc}^{\max})^2\frac{m_W^2}{m_t^2}<m_{lc}^2<(m_{lc}^{\max})^2\frac{1-a}{1+a}, \\[5mm]
\lefteqn{\frac{1+a}{2a(m_{bc}^{\max})^2}\ln\frac{\left(m_{lc}^{\max}\right)^2}{m_{lc}^2}\left(\ln\frac{\left(m_{lc}^{\max}\right)^2}{m_{lc}^2}-\ln\left(2m_{t}^2\right)\right)} \\[5mm]
& \text{for~}
& (m_{lc}^{\max})^2\frac{1-a}{1+a}<m_{lc}^2<(m_{lc}^{\max})^2.
\end{array}
\right.
\label{eq:m_lc_distA}
\end{equation}

\paragraph{Case B}
Here Shape II is excluded, giving 
\begin{equation}
\frac{1}{\Gamma}\frac{\partial\Gamma}{m_{lc}^2}= \left\{ 
\begin{array}{lcl}
\displaystyle \frac{1+a}{2a(m_{bc}^{\max})^2}\ln\frac{1+a}{1-a}\ln\frac{m_t^2}{m_W^2}
& \text{for~}
& 0<m_{lc}^2<(m_{lc}^{\max})^2\frac{1-a}{1+a}\frac{m_W^2}{m_t^2}, \\[5mm]
\lefteqn{\frac{1+a}{2a(m_{bc}^{\max})^2}\left[\ln\frac{1+a}{1-a}\ln\frac{m_t^2}{m_W^2}-\frac{1}{2}\left(\ln\frac{1+a}{1-a}\frac{m_t^2}{m_W^2}\frac{m_{lc}^2}{(m_{lc}^{\max})^2}\right)^2\right]}  \hspace{15mm} \\[5mm]
& \text{for~}
& (m_{lc}^{\max})^2\frac{1-a}{1+a}\frac{m_W^2}{m_t^2}<m_{lc}^2<(m_{lc}^{\max})^2\frac{1-a}{1+a}, \\[5mm]
\lefteqn{\frac{1+a}{2a(m_{bc}^{\max})^2}\ln\frac{m_t^2}{m_W^2}\left(\ln\frac{\left(m_{lc}^{\max}\right)^2}{m_{lc}^2}-\frac{1}{2}\ln\frac{m_t^2}{m_W^2}\right)} \\[5mm]
& \text{for~}
& (m_{lc}^{\max})^2\frac{1-a}{1+a}<m_{lc}^2<(m_{lc}^{\max})^2\frac{m_W^2}{m_t^2}, \\[5mm]
\displaystyle\frac{1+a}{2a(m_{bc}^{\max})^2}\frac{1}{2}\left(\ln\frac{(m_{lc}^{\max})^2}{m_{lc}^2}\right)^2
& \text{for~}
& (m_{lc}^{\max})^2\frac{m_W^2}{m_t^2}<m_{lc}^2<(m_{lc}^{\max})^2.
\end{array}
\right.
\end{equation}


\end{document}